%% file: main.tex
\documentclass[twocolumn,apj,numberedappendix,twocolappendix,iop]{openjournal}

\usepackage[T1]{fontenc}
\usepackage{newtxtext,newtxmath}
\usepackage{xcolor}
\usepackage{textgreek}
\usepackage[utf8]{inputenc}
\usepackage[english]{babel}

\usepackage{color,colortbl}
\definecolor{linkcolor}{rgb}{0.0,0.3,0.5}
\usepackage{tensind}
\tensordelimiter{?}
\DeclareGraphicsExtensions{.bmp,.png,.jpg,.pdf}
\usepackage{verbatim}
\usepackage[normalem]{ulem}
\usepackage{orcidlink}
\usepackage{soul}
\usepackage{natbib}
\usepackage{amsmath}

\usepackage{lineno}

\usepackage{hyperref}
\hypersetup{
    unicode,
    colorlinks=true,
    linkcolor=linkcolor,
    citecolor=linkcolor,
    filecolor=linkcolor,
    urlcolor=linkcolor,
}

\graphicspath{ {./figures/} }

\begin{document}
\title{Mock catalogues for the DESI DR2 Milky Way mass modeling effort}
\input{authors}
\affiliation{The authors' affiliations are listed in Appendix~\ref{app:affiliations}.}

\begin{abstract}
    The Dark Energy Spectroscopic Instrument (DESI) survey is an ongoing campaign to obtain millions of spectra, including for stars within the Milky Way.
    One of DESI's Key Projects for its second data release (DR2) is to measure the mass of the Milky Way using a variety of techniques applied to six-dimensional kinematics of distant halo stars.
    We present mock surveys generated from the Auriga cosmological simulations (AuriDESI) that are used to validate these techniques in several companion studies.
    The AuriDESI tracer catalogues are tailored to reflect key features of the DR2 Key Project dataset, including the selection of RR Lyrae and blue horizontal branch stars, uncertainties on line-of-sight velocities and proper motions, and variation in observational completeness within the survey footprint.
    We discuss how well the resulting AuriDESI tracer catalogues compare to their observed DESI counterparts, including suggesting improvements that can be generalized to similar mock survey efforts within the community.
    The AuriDESI tracer catalogues are publicly available on Zenodo.
\end{abstract}

\maketitle

\section{Introduction}
\label{sec:intro}

\setcounter{footnote}{0} 

It has become clich\'e to say `we are in the era of large surveys' for Galactic astronomy, but it is true nonetheless.
Wide-area digital sky surveys including SDSS \citep{York:2000}, DES \citep{DES:2016}, Pan-STARRS \citep{Chambers:2016}, DELVE \citep{Drlica-Wagner:2021}, and UNIONS \citep{Gwyn:2025} have covered the sky with uniform and deep photometry.
The \textit{Gaia} mission has provided parallax and proper motion measurements for nearly 2 billion stars in the Milky Way \citep{GaiaCollaboration:2018, GaiaCollaboration:2023}.
Spectroscopic surveys including SDSS \citep{York:2000}, DESI \citep{DESI2016a.Science}, and H3 \citep{Conroy:2019} have complemented a subset of these stars with line-of-sight velocities and chemical abundances, which will expand greatly as 4MOST \citep{deJong:2019} and WEAVE \citep{Jin:2024} come online.
Using some combination of the above datasets, Galactic astronomers have discovered more than 60 faint satellite galaxies \citep{Drlica-Wagner:2020, Pace:2025} and more than 100 stellar streams \citep{Grillmair:2016, Ibata:2024, Bonaca:2025}, unravelled the Milky Way's stellar halo into its constituent accretion events \citep{Deason:2024}, and revealed time-dependent disequilibrium in the Milky Way's disk \citep{Antoja:2018, Bennett:2019} and halo \citep{Erkal:2021, Petersen:2021, Vasiliev:2023, Yaaqib:2024, Bystrom:2025, Chandra:2025}.

One quantity of particular interest is the mass (and mass profile) of the Milky Way, which is a key input for mapping satellite counts to dark matter constraints \citep{Newton:2018, Nadler:2020mwsats, Nadler:2021}, predictions for dark matter subhalo interactions with globular cluster streams \citep{Ibata:2002, Johnston:2002,Ngan:2014, Erkal:2016gaps, Nibauer:2025}, and models of the Milky Way potential used for orbit integration \citep{Bovy:2015, McMillan:2017, gala}.
The increase in tracers with available six-dimensional phase-space properties (positions and velocities) have enabled an explosion of measurements for the mass of the Milky Way \citep[e.g.,][]{Patel:2018, Callingham:2019, Watkins:2019, Eadie:2019, Shen:2022h3_mwmass, Ibata:2024, Medina:2025massmodeling} using a variety of tracers (see \citealp{Wang:2020, Bobylev:2023, Bayer:2025} for recent compilations).

Yet even with this dramatic increase in the quantity (number of stars) and quality (information per star) of data, it is not obvious that our measurements are leading to a more \textit{correct} understanding.
In the case of the mass of the Milky Way, several works have highlighted the impact of observational uncertainties and selection functions \citep{Wang:2015, Rehemtulla:2022}, baryonic contraction in the dark matter profile \citep{Cautun:2020, Dado:2026}, and the dynamical influence of the LMC \citep{Garavito-Camargo:2019, Petersen:2020, Erkal:2020, Cunningham:2020, Deason:2021, Brooks:2026} that bias, or at least complicate our interpretations of, the measurements.\footnote{Similar concerns have been raised for mapping halo substructures to individual accreted systems \citep{Thomas:2025, Sante:2026} and the census of faint tidal features \citep{Shipp:2023}.}

Many of these challenges can be addressed by testing measurement methods on adequately sophisticated simulated catalogues, where the answer is known.
Broadly speaking, there have been two approaches to generating mock surveys of Milky Way data.
The first assumes analytic density distributions that are in dynamical equilibrium, most prominently used in the Besan\c{c}on \citep{Robin:2003}, galfast \citep{Juric:2008, Juric:2018}, and TRILEGAL \citep{Girardi:2005} models.
While this approach can be empirically tuned to match observed Milky Way data, it generally oversimplifies the structure of the Galaxy, particularly in the stellar halo where accretion-built substructures (satellites, streams) are modelled as one smooth component.
The second approach is to sample distributions generated by particles in cosmological simulations.
Originally pioneered by \citet{Sharma:2011galaxia}, successive improvements in this approach \citep{Lowing:2015, Thob:2024} have culminated to produce sophisticated mock surveys from flagship cosmological hydrodynamical zoom simulations, including FIRE \citep{Sanderson:2020, Nguyen:2024,Lucey:2026-sdss-mocks} and Auriga \citep{Grand:2018, Kizhuprakkat:2024}.

The Dark Energy Spectroscopic Instrument (DESI) is a state-of-the-art multiplexed spectroscopic instrument operating on the Mayall 4-meter telescope at Kitt Peak National Observatory \citep{DESI2022.KP1.Instr}.
DESI is able to obtain $\sim$5000 spectra at once over a large ($\sim$3~deg) field of view \citep{DESI2016b.Instr, Corrector.Miller.2023, FiberSystem.Poppett.2024}.
While DESI's primary survey objective is to investigate the nature of dark energy by measuring baryonic acoustic oscillations and redshift-space distortions \citep{DESI2024.VII.KP7B, DESI.DR2.DR2}, it has also taken spectra of over 12~million stars in the Milky Way \citep[see Section 5.3 of][]{Koposov:2026}.
The Milky Way Survey (MWS) preferentially targets distant stars in the Galactic halo, including tracers like RR Lyrae (RRL) and blue horizontal branch (BHB) stars for which distances can be precisely measured \citep{Fermani:2013, Catelan:2015, Barbosa:2022, Garofalo:2022}.
One of DESI's Key Projects for its second data release (DR2) is to use halo BHBs and RRLs to measure the mass of the Milky Way, using several methods benchmarked against the same set of mock catalogues generated from cosmological simulations.

In this paper, we present mock catalogues that support this DR2 Key Project effort.
We identify horizontal branch stars from the original AuriDESI target catalogues produced in \citet{Kizhuprakkat:2024}.
We then add mock measurements and survey selection functions that are tailored to the DR2 BHB and RRL catalogues.
The resulting mock tracer catalogues are used in a series of companion papers \citep{Zhou:2026empdf, Medina:2026gme, Valluri:2026nimble} to validate their mass measurement methods, with overall results and interpretation presented in \citet{DESI:2026mwmass}.

The structure of this paper is as follows: in Section~\ref{sec:auridesi} we describe the original AuriDESI catalogues in \citet{Kizhuprakkat:2024}, which are the input for this work.
Section~\ref{sec:tracers} describes the AuriDESI BHB and RRL catalogues that are tailored to the DR2 mass modeling effort.
In Section~\ref{sec:dr2-comparison} we compare these mock tracer catalogues to the DESI DR2 BHB and RRL catalogues.
Finally, Section~\ref{sec:summary} presents a summary and our main conclusions.

\section{AuriDESI framework} \label{sec:auridesi}

The mocks presented in this work are largely building off of previous AuriDESI catalogues \citep{Kizhuprakkat:2024}, which are themselves developed from the AuriGaia mocks presented in \citet{Grand:2018}.
In this Section we summarize the procedure to generate the \textit{photometric target catalogues} described in \citet{Kizhuprakkat:2024}, which are the input catalogues used in this work.
Our adaptations to these catalogues that are specific for the DR2 Key Project, including selecting horizontal branch stars and applying observational uncertainties, are detailed in Section~\ref{sec:tracers}.

\subsection{Auriga simulations}

\begin{figure*}
    \centering
    \includegraphics[width=\linewidth]{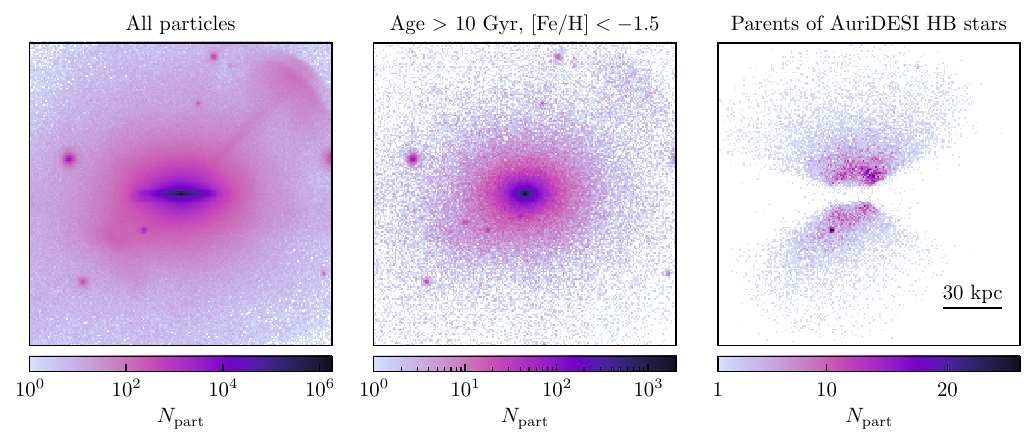}
    \caption{Spatial distribution of all star particles in Au-27 (left), those that are old and metal-poor (middle), and those that spawn a BHB or RRL in the 30~deg viewing mock DESI survey.
    The spatial extent and bin size are the same in each panel.
    A physical indicator of 30~kpc is shown in the right-most panel.
    }
    \label{fig:spatial-distribution-particles}
\end{figure*}

Our mock catalogues are based on simulated Milky Way-mass galaxies from the Auriga project \citep{Grand:2017, Grand:2024}.
The haloes were selected from the dark-matter-only 100 Mpc box from the EAGLE project \citep{Schaye:2015} to be isolated and have a halo mass\footnote{This virial mass is the mass enclosed in a sphere in which the mean matter density is 200 times the critical density $\rho_\text{crit} = 3H^2(z) / 8\pi G$.} $M_\text{200c} = 1 - 2 \times 10^{12}$~M$_\odot$
The simulations are described in further detail in \citet{Grand:2017, Grand:2024}.
An example simulated halo is presented in Figure~\ref{fig:spatial-distribution-particles}.

They were resimulated with the moving-mesh code \textsc{Arepo} \citep{Springel:2010, Pakmor:2016} from zoom-in initial conditions generated using \textsc{Panphasia} \citep{Jenkins:2013, Jenkins:2024}.
The Auriga galaxy formation model includes primordial and metal-line cooling \citep{Vogelsberger:2013}, star formation and stellar feedback \citep{Springel:2003, Vogelsberger:2013}, a spatially-uniform redshift-dependent UV background for reionization \citep{Faucher-Giguere:2009}, black hole seeding, accretion, and feedback \citep{Springel:2005feedback, Marinacci:2014, Grand:2017}, and magnetic fields \citep{Pakmor:2017}.
The Auriga project includes simulations at several numerical resolution levels; our mocks are based on `level 3' runs with baryonic element (dark matter particle) masses of 6.7(36)$\times 10^3$~M$_\odot$ and a minimum softening length of 188 pc.
There are six Milky Way-mass haloes simulated as this resolution, which we designate as Au-(6, 16, 21, 23, 24, 27).

The Auriga model produces spiral disc galaxies that are broadly consistent with a number of observations including stellar masses, sizes, and rotation curves \citep{Grand:2017}, H~\textsc{i} gas distributions \citep{Marinacci:2017}, stellar disc warps \citep{Gomez:2017}, stellar bars \citep{Fragkoudi:2020, Fragkoudi:2025}, and magnetic fields \citep{Pakmor:2017}.
In particular for our modelling efforts, the simulations naturally produce accreted substructures including satellite galaxies \citep{Simpson:2018}, stellar streams \citep{Riley:2025, Shipp:2025}, and stellar haloes \citep{Monachesi:2019}.
The accreted components of Auriga galaxies have been studied in a variety of relevant scientific contexts, including inferring the mass of the Milky Way \citep{Callingham:2019, Deason:2019} and M31 \citep{Gherghinescu:2024}, identifying and distinguishing accretion events in the stellar halo \citep{Thomas:2025, Sante:2026}, characterizing dominant contributors that are analogues of the Gaia-Sausage-Enceladus (GSE) merger \citep{Fattahi:2019, Carrillo:2024, Carrillo:2026} and the overall build-up of the accreted stellar halo \citep{Fattahi:2020, Deason:2023, Orkney:2023, Pu:2025}.

\subsection{Generating stars} \label{sec:generating-stars}

Cosmological simulations like Auriga produce star particles that each represent an initial mass function (IMF)-averaged simple stellar population (SSP) of a given age and metallicity.
For a survey of resolved stars like DESI, producing a mock catalogue from such simulations requires transforming these star particles into individual stars.
These stars can then be assigned stellar parameters (e.g. $T_\text{eff}$, $\log g$, $grz$ photometry) that are convolved with survey properties like magnitude-dependent uncertainties and selection functions.

The key ingredients for turning star particles into stars for AuriDESI are as follows:
\begin{itemize}
    \item Coordinate system: each galaxy is centered and rotated such that the host's stellar disc aligns with the Cartesian $z=0$ plane. Four choices for the solar position are spread at equidistant azimuthal angles relative to the bar\footnote{Labelled 30, 120, 210, and 300~deg in the output files.}, with the primary reference angle trailing the bar by 30~$\deg$ \citep{Wegg:2015, Bland-Hawthorn:2016}.
    \item IMF sampling: each star particle's initial mass is split according to a \citet{Chabrier:2003} IMF\footnote{Specifically the coefficients and functional form described in Table~1 of \citet{Chabrier:2003}, which is fit to individual stars in the Galactic disk.} sampled from 0.08 to 120~M$_\odot$. For a given SSP, stars are only generated within a mass range set by the limiting magnitude of the survey at the faint end and the maximum stellar mass for a given age at the bright end.
    \item Stellar parameters: once an age, metallicity, and initial mass are determined for each star, stellar parameters and photometry are assigned using a grid of PARSEC v1.2S isochrones \citep{Bressan:2012, Chen:2014, Chen:2015, Tang:2014}. These isochrones do not include improved treatment for thermally pulsing asymptotic giant branch stars \citep[\textsc{colibri};][]{Marigo:2013}. We do not interpolate \textit{between} or \textit{along} isochrones; stars are assigned to the isochrone in the grid with the nearest age and metallicity and to the point along that isochrone with the nearest initial mass.
    \item Phase-space sampling: to avoid discrete `clumps' of stars at the exact location of their parent star particle, stars are assigned phase-space coordinates according to the procedure described in \citet{Lowing:2015}. This process estimates a 6D phase-space kernel for each particle that is locally smoothed and re-scaled to preserve local velocity anisotropies. To avoid `cross-talk' between overlapping phase-space structures when determining the neighboring particles used to define the kernel, stars are grouped based on their origin. Accreted star particles are grouped based on the galaxy they formed in, while \textit{in situ} stars are treated as one group.
\end{itemize}

\subsection{Parent survey selection functions} \label{sec:parent-selection-functions}

Both AuriGaia and AuriDESI applied selections in magnitude and sky position relevant to their respective surveys.
The union of these two is the following:
\begin{itemize}
    \item Sky position: Galactic latitude $|b| > 20\deg$ and falling within the DESI footprint, defined by the union of bright-time observing tiles used in \citet{Kizhuprakkat:2024}
    \item Magnitude: $16 < V < 20$,\footnote{We note that this selection in $V$-band is inherited from AuriGaia \citep{Grand:2018}.} $16 < r < 20$
\end{itemize}
The AuriGaia catalogues do not include stars fainter than $V=16$ for $|b| < 20\deg$, however this is outside of the DESI footprint.
The AuriDESI target catalogues also assigned each star to one of several targeting classes that are used to prioritise spectroscopic targeting in the survey (see Table~1 of \citealp{Kizhuprakkat:2024} and \citealp{Cooper:2023}).
However, this priority ordering does not affect whether a star is included in the catalogue, and it is ignored when constructing the tracer catalogues in this work (Section~\ref{sec:tracers}).

After these procedures to generate stars and apply simple survey selection functions, the result is a magnitude-limited, complete catalogue of stars that fall within the DESI footprint.
There are four catalogues (corresponding to the observer locations) for each of six Auriga haloes, resulting in 24 mock catalogues in total.

\section{Mock tracer catalogues} \label{sec:tracers}

In this Section we describe our procedure to construct the AuriDESI \textit{tracer catalogues} from the photometric target catalogues described in Section~\ref{sec:auridesi}.
These are analogous to the BHB and RRL catalogues used in the DR2 Key Project.
We begin by briefly describing these catalogues and refer to Sections~2.2 and 2.3 of \citet{DESI:2026mwmass} for further detail.

\subsection{The DESI DR2 tracer catalogues} \label{sec:real-catalogues}

The DESI MWS \citep{Cooper:2023} shares the focal plane with the DESI Bright Galaxy Survey (BGS; \citealp{BGS.TS.Hahn.2023}) during bright observing conditions, utilizing otherwise unassigned fibers to observe stars in an efficient manner \citep{SurveyOps.Schlafly.2023}.
The main MWS program targets stars with $16 < r < 19$, with additional surveys and programs for brighter stars \citep{Dey:2026}, the Andromeda Galaxy \citep{Dey:2023}, dwarf galaxies and stellar streams \citep[e.g.][]{Ding:2025, Bystrom:2026}.
A subset of rare stars, including BHBs and RRLs, are given the highest targeting priority during fiber assignment (see \citealp{Cooper:2023} for details).
Nightly observations are reduced by an automated pipeline \citep{Spectro.Pipeline.Guy.2023} and periodically collected and uniformly re-processed for public release, the latest of which \citep[DR1;][]{DESI2024.I.DR1} included spectra for more than four million stars \citep{Koposov:2026}.
The DR2 dataset covers the first three years of DESI observations, contains spectra for more than 12 million stars, and is scheduled for release in early 2027.

The DR2 BHB catalogue is constructed in the same manner as in \citet{Bystrom:2025} for DR1.
BHBs are targeted using photometry and astrometry, with contaminants including quasars, blue straggler stars, and RRLs removed using DESI spectral information.
Distances are derived using the absolute magnitude-colour relation in Equation~6 from \citet{Bystrom:2025}, while line-of-sight velocities and iron abundances are derived from the RVS pipeline \citep{Koposov2019rvspecfit, Koposov:2026}.

The DR2 RRL catalogue is an expansion of the DR1 catalogue presented in \citet{Medina:2025paper1}.
Targets are identified from the \textit{Gaia} RRL catalogue \citep{Clementini2018}.
Distances are derived using the absolute magnitude-metallicity relation of \citet{Garofalo:2022}, based on dereddened Gaia $G$-band photometry and iron abundances derived from DESI spectra and averaged across epochs.
Systemic line-of-sight velocities are determined empirically using a Bayesian radial velocity curve methodology described in \citep{Medina:2025paper1}.
This technique uses \textit{Gaia} DR3 periods and an empirically-determined radial velocity phase curve to simultaneously fit all DESI epochs for a particular RRL.
The DR2 RRL catalogue will be presented in an upcoming paper (Medina et al.~in preparation) outside of the Key Project effort.

From these input catalogues, we apply several selections that are applied the same way to both tracers.
While both catalogues contain stars observed during bright and dark time, for the Key Project we restrict to only use bright time observations taken as part of the main survey to have a more straightforward selection function.
Both catalogues are augmented with \textit{Gaia}~DR3 proper motion measurements.
They are also cross-matched to external catalogues of member stars in satellite galaxies \citep{Pace:2022} and globular clusters \citep{Vasiliev:2021gcs_edr3} to remove known halo substructures.
We also remove stars from the Sagittarius stellar stream based on sky position and heliocentric distance.
After all of these selections, there are 4838 RRLs and 5198 BHBs in total, extending to $\sim$90~kpc from the centre of the Galaxy.

\subsection{Defining a BHB and RRL in PARSEC} \label{sec:tracer-definition}

\begin{figure*}
    \centering
    \includegraphics[width=\linewidth]{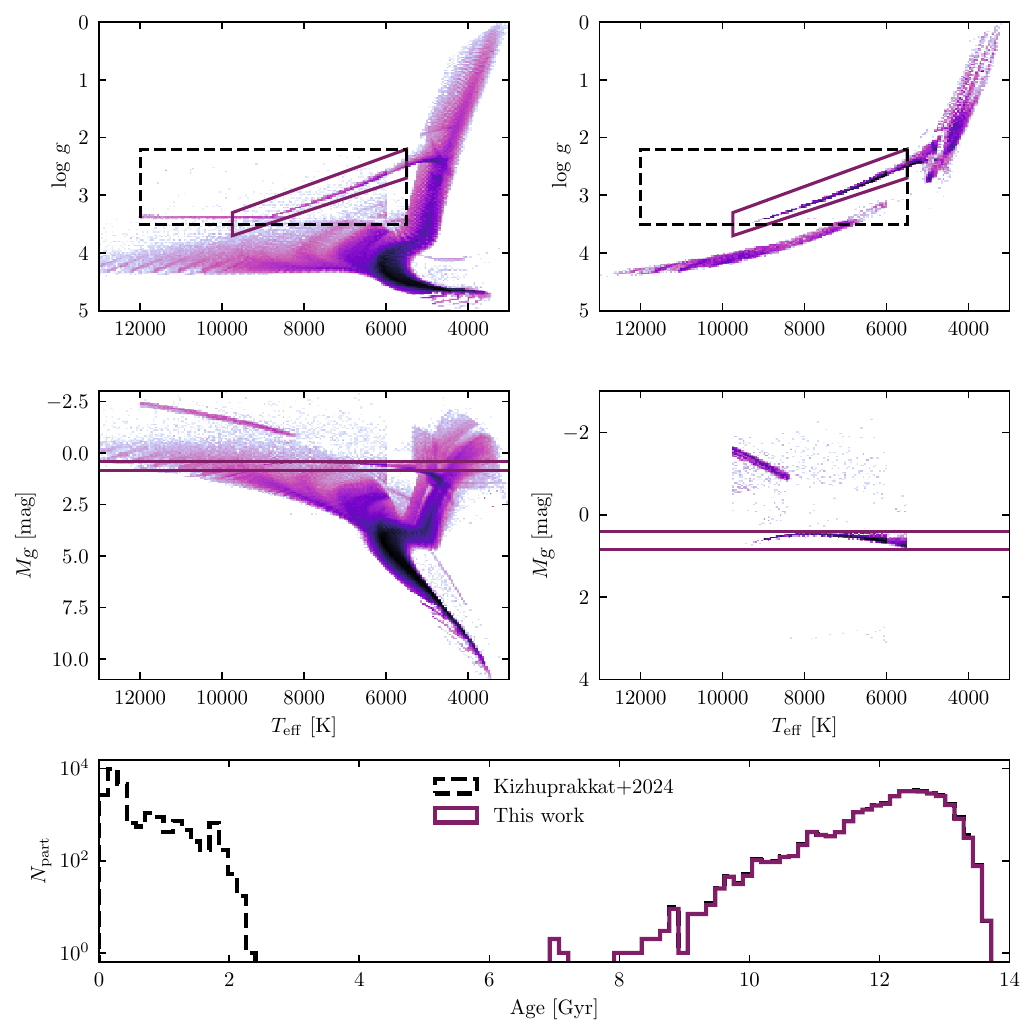}
    \caption{Horizontal branch definition and comparison with \citet{Kizhuprakkat:2024}.
    Top and middle left: distribution of all stars from the Au-21 30~deg viewing mock catalogue in $\log g$, $T_\text{eff}$, and $M_g$.
    The black dashed box is the horizontal branch selection used in \citep{Kizhuprakkat:2024}, while the red boxes are the selections used in this work.
    Top and middle right: distributions in the same planes after applying only absolute magnitude (top) or $\log g - T_\text{eff}$ selections (middle) from this work.
    This shows the need to require a selection in all three parameters to efficiently extract horizontal branch stars.
    Bottom: distribution of ages for stars that satisfy the horizontal branch selection in \citet{Kizhuprakkat:2024} (black) and this work (red).
    }
    \label{fig:hb-selection}
\end{figure*}

The PARSEC isochrones used to generate mock stars in AuriDESI include a simple model for the horizontal branch.
We select stars that fall along this horizontal branch using the following criteria:
\begin{itemize}
    \item Temperature: $5500 \text{~K} < T_\text{eff} < 9750 \text{~K}$
    \item Surface gravity: $0.7765 + T_\text{eff}/3863 < \log g < 1.4059 + T_\text{eff}/4250$
    \item Absolute magnitude: $0.4 < M_g < 0.85$
    \item Age: $t_\text{form} > 8$~Gyr
\end{itemize}
These selection criteria are illustrated in Figure~\ref{fig:hb-selection}.
The full distribution of stars in $\log g$, $T_\text{eff}$, and $M_g$ from the Au-21 30~deg target catalogue is shown in the left panels.
The right panels show stars that pass either the Kiel space selection or the absolute magnitude selection.
The combination of these selections is required to select horizontal branch stars efficiently.

We also show the definition of horizontal branch stars used in \citet{Kizhuprakkat:2024}, which was a selection in Kiel space ($2.2 < \log g < 3.5$ and $5500 < T_\text{eff} < 12000$) with no selection in absolute magnitude.
These broader criteria included stars outside of the horizontal branch - in particular, they include a substantial population of young ($<$2~Gyr) blue loop stars \citep{Tang:2014} that were often located in massive, star forming satellites within the DESI footprint.
The amount of contamination from this population depended on the particular halo and viewing within a halo (satellites, or a subset of the stars within them, can move outside the footprint or magnitude limit), but was as high as $\sim$43~per~cent for Au-21.
We note that an over-representation of young stars in a priority target class has non-trivial downstream effects for the `spectroscopic' and `re-sampled' catalogues presented in \citet{Kizhuprakkat:2024}.
This should be addressed in future versions of AuriDESI that produce catalogues including all stellar populations accessible to DESI.

Once we have selected horizontal branch stars, we need to determine which of these stars are RRLs and which are BHBs.
In reality, this is a complex boundary that depends on internal structural properties of stars on the instability strip, which is not captured in the PARSEC models.
For the DR2 RRL and BHB catalogues, stars hotter than $T_\text{eff}\sim 7375$~K tend to be BHBs while cooler stars tend to be RRLs, though this boundary is somewhat porous due to abrupt changes in surface temperature for RRLs \citep[see][]{Medina:2025paper1, Medina:2025paper2}.
For simplicity, we adopt a separation boundary at $T_\text{eff} = 7000$~K; stars that are hotter than this threshold labelled as BHBs, while stars that are cooler are labelled as RRLs.

\begin{figure}
    \centering
    \includegraphics[width=\linewidth]{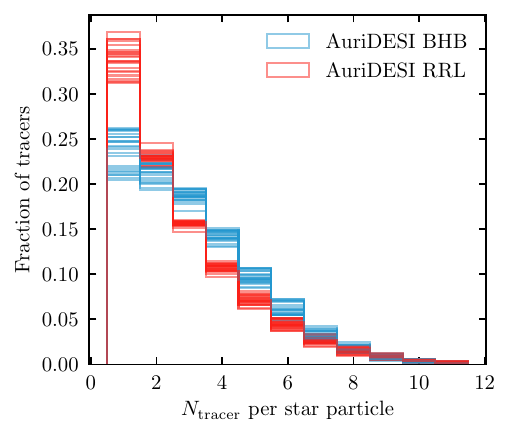}
    \caption{Number of AuriDESI BHBs (blue) or RRLs (red) produced per Auriga star particle.
    Each of the 24 mocks (6 haloes with 4 viewing angles per halo) is represented as a thin line.
    }
    \label{fig:tracers-per-particle}
\end{figure}

For our adopted definitions, star particles in Auriga typically generate multiple horizontal branch stars.
Figure~\ref{fig:tracers-per-particle} shows the distribution of BHBs or RRLs produced for a unique star particle.
On average, there are $\sim$2.8~RRLs per particle and $\sim$3.1~BHBs per particle.
Often the same particle can generate both types of tracers; for example, typically $\sim$55~per~cent of BHB-generating particles also produce at least one RRL, while $\sim$30~per~cent of RRL-producing particles also produce at least one BHB.
We note that the $N_\text{RRL}-M_V$ relationship for Local Group satellite galaxies \citep{Martinez-Vazquez:2019, Martinez-Vazquez:2023} implies a rate of $\sim$5.4 RRLs per particle at Auriga level 3 resolution and adopting $M_\star / L_V = 2$, though this is substantially higher than rates for old star clusters in the LMC \citep{Sarbadhicary:2021, Cuevas-Otahola:2025} that imply $\sim$0.15 RRLs per particle.

Finally, we note that these selections are applied using true properties (i.e., before convolving with uncertainties).
While it can be desirable to derive a catalogue using error-convolved values that closely mimics the survey procedure, the primary goals such an excercise would be to capture (1)~real BHBs and RRLs leaving the selection and (2)~contaminants from other sources passing these criteria.
We found that for our specific tracer population, uncertainties on $T_\text{eff}$, $\log g$, and apparent $grz$-magnitudes are small enough that real horizontal branch stars rarely leave reasonable observational selection windows.
In addition, the primary sources of contamination are either photometric targets that would be removed when constructing the final catalogues (e.g., active galactic nuclei or blue straggler stars that mimic photometry of BHBs) or are not adequately modeled by PARSEC and therefore not captured in the AuriDESI framework (e.g., eclipsing binaries or other sources with variable light curves that mimic RRL).
Since our priority for these catalogues is to assess the impact of survey properties (selection function, uncertainties) and galaxy disequilibrium on Milky Way mass estimates, we find using true tracer properties reasonable for this step.

\subsection{Selection function} \label{sec:desi-selfunc}

The selection function for DR2 BHBs and RRLs has two components: (1) overall DESI survey properties (e.g., footprint, magnitude limit) and (2) variations in completeness along the sky that are specific to DR2.
The first component was described in Section \ref{sec:parent-selection-functions}, but in summary, BHBs and RRLs are only included as photometric targets if they are within the DESI footprint and brighter than $r=20$.
This component of the selection function is naturally accounted for in the steps to construct the mock catalogues described to this point.

The second component of the survey selection function accounts for on-sky variation within the DESI footprint.
DESI is a eight-year spectroscopic campaign that remains ongoing, with DR2 corresponding to the first three years of observations.
By the end of those observations, some regions of sky (or `tiles') had been observed up to five times, while others had been observed only once.

The details of how this on-sky variation is characterized are described in Section 2.4 of \citet{DESI:2026mwmass}.
The relevant outcome for constructing the mocks is a mapping from on-sky position (and for RRLs, apparent magnitude in the \textit{Gaia} $G$-band) to a detection probability $p_\text{obs}$.
For RRLs, $p_\text{obs}$ captures effects from both the DESI survey and \textit{Gaia} DR2, which is the input catalogue used for targeting.
We use the \textit{Gaia} DR2 RRL selection function empirically determined by \citet{Mateu:2020} for the \textit{Gaia} component.
For BHBs, $p_\text{obs}$ only explicitly captures effects from the DESI survey, while the input target catalogues are assumed to be 100~per~cent complete.
We note that the magnitude limit of the DR2 BHB catalogue ($r=20$) is considerably brighter than the equivalent for the Legacy Survey ($r=23.9$) and \textit{Gaia} ($G=20.7$) used for targeting and proper motions, respectively.

\begin{figure*}
    \centering
    \includegraphics[width=\linewidth]{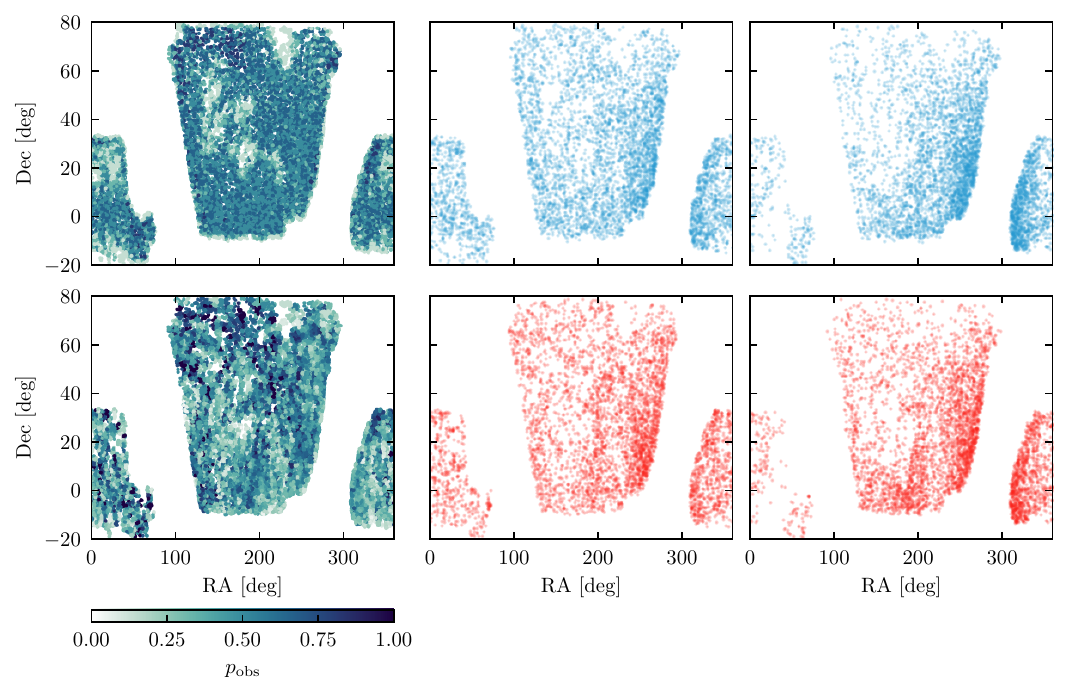}
    \caption{On-sky distribution of BHBs (top panels) and RRLs (bottom panels). Left: the photometrically-complete sample of AuriDESI BHBs and RRLs in the 30~deg viewing of Au-6, coloured by completeness $p_\text{obs}$. Middle: AuriDESI tracers remaining after all downsampling procedures. Right: observed DESI DR2 tracers.
    Note that the sharp underdensity in DR2 RRLs within $150^\circ < \text{RA} < 200^\circ$ and $0^\circ < \text{Dec} < 40^\circ$ is a combined effect of the \textit{Gaia} selection function and Sagittarius stream removal.}
    \label{fig:spatial-pobs-error}
\end{figure*}

The variation in $p_\text{obs}$ across the sky is shown in the left-hand panels of Figure~\ref{fig:spatial-pobs-error}.
We show all AuriDESI BHBs and RRLs within the DESI footprint, coloured by their $p_\text{obs}$ values.
Most of these locations are at declinations $\delta \gtrsim 80$~deg or $\delta \lesssim -15$~deg, where azimuthal conditions are not favorable.
The difference in smoothness between BHB and RRL are due to the techniques when computing the DESI completeness (see \citealp{DESI:2026mwmass}).

Once we have assigned $p_\text{obs}$ to each individual star using these selection functions, we apply this on-sky variation in completeness to the mocks.
For each star, we generate a random number $X$ sampled from a uniform distribution $\mathcal{U}(0,1)$.
If $X < p_\text{obs}$, that star is labelled as `observed' in our mock catalogues.
Since some methods require repeating this excercise several times for calibration \citep{Zhou:2026empdf}, in the final catalogues we retain stars that are not observed and provide a column that captures if stars were determined to be `observed' at this step.

After downsampling the catalogue according to $p_\text{obs}$, there are still many more AuriDESI BHBs ($\sim$9,000--15,000) and RRLs ($\sim$10,000--18,000) than there are in the DESI DR2 BHB ($\sim$5500) and RRL ($\sim$4800) catalogues.
This primarily arises from the fact that the Auriga galaxies that AuriDESI is based on have more massive stellar haloes than the Milky Way's (e.g., Figures 17 and 18 of \citealp{Monachesi:2019}), resulting in more BHB and RRL stars in both the target catalogues and after applying selection functions, though it is also sensitive to our adopted stellar evolution model (see Section~\ref{sec:dr2-comparison}).
To further downsample the catalogue for case studies where the number of tracers is important, we provide a randomly generated integer ID (from 1 to the number of tracers in the catalogue) for both the full catalogue and the subsample that is `observed'.
For the DR2 Key Project modeling, we downsample all catalogues to include exactly 4850 RRLs and 5550 BHBs \textit{after} removing halo substructures (see Section~\ref{sec:removing-halo-substructures}).
Equivalent numbers for the DESI DR2 catalogues are 4838 RRLs and 5198 BHBs (see Section~\ref{sec:real-catalogues}).

\subsection{Error model} \label{sec:error-model}

\begin{figure*}
    \centering
    \includegraphics[width=\linewidth]{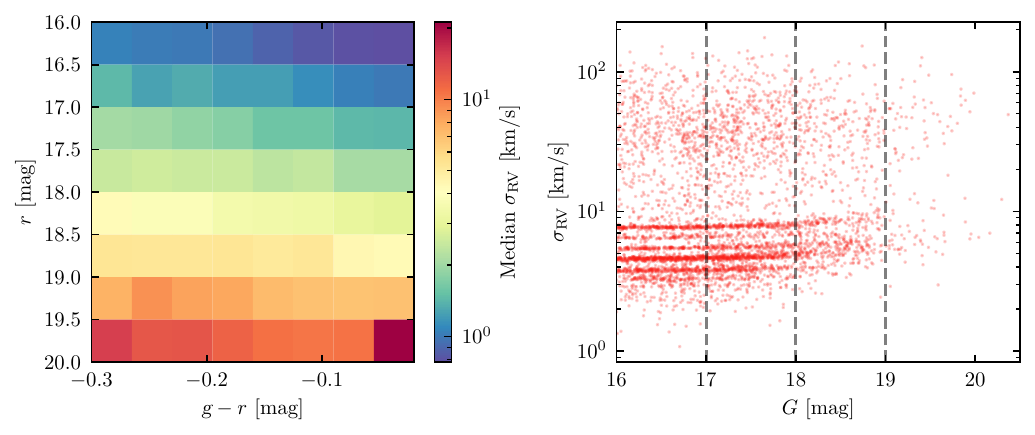}
    \caption{Error model for BHB and RRL tracers.
    Left panel: median line-of-sight velocity uncertainty as a function of $g-r$ and $r$ for the DESI DR2 BHBs.
    AuriDESI BHBs are assigned the median uncertainty from the corresponding bin.
    Right panel: line-of-sight velocity uncertainty as a function of $G$-band magnitude for RRLs in DESI DR2.
    AuriDESI RRLs are binned in $G$ according to the dashed vertical lines, then assigned an uncertainty randomly sampled from the real DR2 RRL uncertainties in that same bin.
    }
    \label{fig:vrad-error-model}
\end{figure*}

We convolve observational quantities relevant to DR2 Key Project with uncertainties that mimic key aspects of the real survey.
For all quantities, we sample a Gaussian with a mean of zero and a width corresponding to the assigned uncertainty, then add this to the true value.
We note that no uncertainties are applied to the photometry (see final paragraph of this subsection) and uncertainties that depend on photometric quantities use the corresponding true value.

Astrometric quantities (parallax, proper motions) are convolved with uncertainties using the \texttt{pygaia} package.\footnote{\url{https://github.com/agabrown/PyGaia}}
These uncertainties, specifically corresponding to DR3, are calibrated to the mean performance of the \textit{Gaia} survey and only depend on apparent $G$-band magnitude:
\begin{align}
\sigma_X &= c_X \sigma_\varpi, \quad \sigma_\varpi = \sqrt{40 + 800z + 30z^2} \\
\log_{10}z &= 0.4(\text{max}[G, G_\text{bright}] - 15)
\end{align}
where $G_\text{bright} = 13$, $c_{\mu_{\alpha*}} = 1.03$, and $c_{\mu_{\delta}} = 0.89$.

Distance uncertainties are assumed to be 5~per~cent for both BHBs and RRLs.
While the DESI DR2 BHB catalogue does not provide uncertainties to individual stars, a sample of BHBs cross-matched to dwarf galaxies and globular clusters with literature distances \citep{Vasiliev:2021gcs_edr3, Pace:2022} had a residual scatter of 5.7~per~cent on average.
The DESI DR2 RRL catalogue provides uncertainties for individual stars that are derived by \citet{Garofalo:2022}, with a typical distance uncertainty of $\sim$2~per~cent.
For simplicity, we adopt a uniform 5~per~cent distance uncertainty for both tracers.

The uncertainties on line-of-sight velocities are customized to DESI DR2 and are developed differently for BHBs and RRLs.
For BHBs, DESI performance is a function of colour and magnitude, similar to uncertainty models developed in prior work (e.g., Figure~10 in \citealt{Cooper:2023} and Figure~6 in \citealt{Kizhuprakkat:2024}).\footnote{We note that the uncertainties in these earlier models, which were constructed in $g-r$ vs.~$r$ for all stars in the MWS, were somewhat exaggerated, particularly for blue stars. This is because they included white dwarfs that have higher \texttt{VRAD} errors from the RVS pipeline at a given colour and magnitude. Future versions of these models should remove white dwarfs prior to computing survey performance, in addition to other improvements (e.g., binning in [Fe/H] vs.~$z$, which captures overall survey properties for a wider selection of stars than those examined here).}
We bin the DR2 BHB catalogue in $g-r$ vs.~$r$ and compute the median line-of-sight velocity uncertainty in each bin.
In Figure~\ref{fig:vrad-error-model} we show both the uncertainties as a function of $g-r$ vs.~$r$ and this resulting error model.
AuriDESI BHBs are binned in the same fashion and assigned the median uncertainty shown in Figure~\ref{fig:vrad-error-model}.

The uncertainties for systemic line-of-sight velocities of the RRL catalogue are more complicated to model.
Since RRLs are pulsating variable stars, their line-of-sight velocity measurements vary substantially with time.
\citet{Medina:2025paper1} modelled this variation empirically based on DESI data alone, using periods determined from \textit{Gaia}.
The resulting uncertainty on the systemic line-of-sight velocity for an individual RRL primarily depends on the number of epochs and the phase of the RRL at the time of those observations, due to abrupt variations in atmospheric kinematics between minimum and maximum light.

We show the distribution of these uncertainties in Figure~\ref{fig:vrad-error-model} as a function of dereddened magnitude in the Gaia $G$-band.
For brighter stars, this distribution has several notable peaks between 3 and 8~km~s$^{-1}$, along with a substantial tail greater than 11~km~s$^{-1}$.
Once again, the high-error tail is predominantly composed of RRLs with some combination of (a) fewer DESI epochs and (b) epochs that occur near maximum light.
For fainter stars, the peaks become less prominent.
In order to capture these features, we first bin both DESI and AuriDESI tracers in $G$-band magnitude (brighter than $G=17$, 17 to 18, 18 to 19, and fainter than $G=19$).
Then we sample the distribution of DESI line-of-sight velocity uncertainties directly (i.e., drawing with replacement) for a given bin and assign those uncertainties to the mock RRLs binned in the same manner.

Since these mock catalogues are tailored to the DR2 Key Project, there are several observational quantities that might be convolved with errors for a more general effort (as in \citealt{Grand:2018} and \citealt{Sanderson:2020} for \textit{Gaia}) that we do not.
The modeling methods in the Key Project do not explicitly depend on $grz$ photometry that has small uncertainties ($\lesssim$0.003 mag for Legacy Survey imaging at $r=20$), nor do they depend on spectrosopic quantities like $\log g$ and $T_\text{eff}$ beyond filtering contamination (see the end of Section~\ref{sec:tracer-definition}).
The astrometric uncertainty for the on-sky location (RA, Dec) of \textit{Gaia} sources is negligible as well.

\subsection{Removing halo substructures} \label{sec:removing-halo-substructures}

Tracers that are members of coherent halo substructures are known to bias techniques to model the mass of the Milky Way \citep{Han:2016, Wang:2017dynamicalmodelling, Wang:2018, Shen:2022h3_mwmass, Medina:2025massmodeling}.
Both the DR2 BHB and RRL catalogues identify stars for removal that are members of known globular clusters, dwarf galaxies, and the prominent Sagittarius stellar stream.
This is based on the 3-D spatial positions of the stars relative to literature catalogues, as described in Sections~2.2 and 2.3 of \citet{DESI:2026mwmass}.
In a similar manner, we provide substructure flags for our AuriDESI mock catalogues.

Satellites in Auriga are identified using the SUBFIND algorithm, which identifies self-bound substructures within friends-of-friends (FOF) groups.
These (sub-)haloes are tracked over time using merger trees constructed by LHaloTree \citep{Springel:2005}.
AuriDESI stars are assigned the same status (e.g., bound to a subhalo, accreted/in-situ) as their parent star particle, as described in \citet{Kizhuprakkat:2024}.
We provide several columns that encode this information, detailed in Appendix~\ref{app:data-model}.
With this information, stars that are bound to satellites at the present day are easily removed.

We also flag AuriDESI stars that originate from star particles in massive stellar streams.
\citet{Riley:2025} and \citet{Shipp:2025} characterized all accreted systems in the Auriga simulations and classified their present day morphology as `intact', `stream', or `phase-mixed'.
Systems labelled as `streams' have lost at least 3~per~cent of their stars and are dynamically cold (low median local velocity dispersion at a given stellar mass, see \citealp{Riley:2025} for details).

\begin{figure*}
    \centering
    \includegraphics[width=\linewidth]{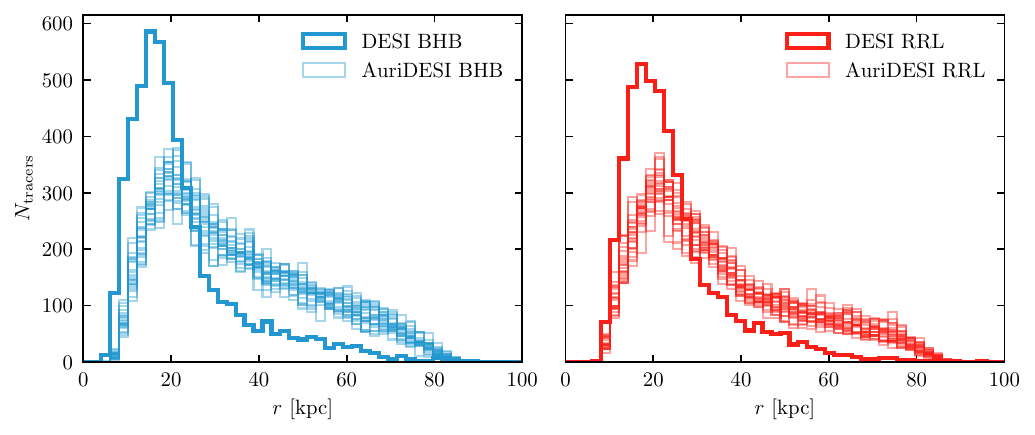}
    \caption{Distances for BHB and RRL tracers in DESI and AuriDESI, where each AuriDESI mock is represented as a thin line (24 mocks in total).
    The mock distributions are after all downsampling procedures, including to match the number of tracers in the DESI DR2 catalogues.
    AuriDESI BHBs and RRLs are less centrally concentrated around the host galaxy than the real DESI data.
    }
    \label{fig:distance-comparison}
\end{figure*}

Stars are flagged as members of stellar streams if their parent star particle belongs to a stream identified in \citet{Riley:2025} above a certain stellar mass threshold.
One complication is that the stellar mass-halo mass relation in the Auriga simulations is high relative to both observational constraints and other comparable simulations.
We provide flags for two mass thresholds:
\begin{itemize}
    \item $M_\star > 10^8$~M$_\odot$: the summed stellar mass of Sagittarius's bound component and tidal tails \citep{Niederste-Ostholt:2010}
    \item $M_\star > 2\times10^9$~M$_\odot$: the stellar mass that results from taking the $10^8$~M$_\odot$ from above, converting it to a (peak) halo mass \citep{Behroozi:2013smhmr}, and converting this halo mass to a stellar mass using the Auriga stellar mass-halo mass relation
\end{itemize}
The DR2 Key Project adopted the $M_\star > 10^8$~M$_\odot$ threshold when testing mass modelling techniques.

We note that the particle lists that identify accreted star particles and assign them to a system are not the same between the AuriGaia catalogues \citep{Grand:2018} and those used to identify streams \citep{Riley:2025}.
This means a star particle could be considered a member of one system during the phase-space sampling described in Section~\ref{sec:generating-stars} and a different system when identifying stellar streams.
For each massive stellar stream, we perform a cross-match between star particles in the \citet{Riley:2025} catalogues and star particles that produce stars in the \citet{Kizhuprakkat:2024} spectroscopic target catalogues.
In each case, the particles are predominantly assigned to one system in the AuriGaia catalogues, with $\lesssim$0.5~per~cent of particles assigned to others.
We then label all stars generated with the dominant AuriGaia progenitor system as belonging to a massive stream.
Removing these stars results in ``smooth'' stellar haloes that are that are void of cold substructures massive enough to impact mass modeling efforts.
We note that phase-mixed accreted substructures, as well as lower mass streams, are retained in the catalogues.

\section{Comparison to DR2 data} \label{sec:dr2-comparison}

In this Section we assess how well the AuriDESI BHB and RRL catalogues capture key features of their observed DESI counterparts.
We focus on phase-space coordinates (Section~\ref{sec:phase-space-comparison}), ages and metallicities (Section~\ref{sec:age-feh-comparison}), and relative numbers of BHB and RRL (Section~\ref{sec:bhb-rrl-ratio-comparison}).

\subsection{Phase-space coordinates} \label{sec:phase-space-comparison}

In Figure~\ref{fig:distance-comparison} we show the distribution of tracers in Galactocentric distance.
The distribution for both the mocks and DESI data begins near 10~kpc, peaks near 20~kpc, and stops near 90~kpc.
However, the real DESI tracers are more centrally concentrated about the host; $\sim$80~per~cent fall within 30~kpc for the real data, while the mocks only contain $\sim$50~per~cent within the same range.
The discrepancy gets stronger with distance, with the mocks having $\sim$20~per~cent of stars beyond 50~kpc and $\sim$0.5~per~cent beyond 80~kpc, while the real data have $\sim$5~per~cent and $\sim$0.1~per~cent, respectively.
While the accretion histories vary substantially between different Auriga haloes \citep{Monachesi:2019, Orkney:2023, Kizhuprakkat:2024, Riley:2025, Pu:2025, Thomas:2025}, the distance distribution of AuriDESI BHBs and RRLs is remarkably consistent across different haloes and viewings.

We note that we are showing the AuriDESI tracer distribution after re-sampling to match the number of tracers in the DR2 catalogues, so this should be considered a comparison between the relative contribution in a particular radial bin.
This overabundance of total mock stars (after accounting for footprint and completeness) is primarily due to the high masses for Auriga stellar haloes.
Since Auriga has a higher stellar mass - halo mass relation than both observations and comparable simulations, this results in more massive stellar haloes than typically predicted in other theoretical models.

The discrepancy in the \textit{normalised} radial distribution of tracers has two main causes.
The first is that the slope of the radial density profile of old ($t_\text{form} > 10$~Gyr) and metal-poor ($\mathrm{[Fe/H]} < -1.5$) stars in Auriga may be shallower than in the Milky Way.
A shallower profile would result in a larger fraction of tracers further away from the center of the host galaxy.
While \citet{Monachesi:2019} studied radial profiles of Auriga galaxies and found a general agreement with available observations, these studies analysed the total stellar distribution instead of old and metal-poor that we focus on here.
Recent work by \citet{Lucey:2026-missing-halo} shows that the slopes and breaks in the density profile of Auriga provide a reasonable match to the Milky Way, particularly if only accreted, only old, or only metal-poor stars are considered.
We also note that \citet{Thomas:2021} compared stellar density profiles of PAndAS-like mocks of the Auriga simulations to M31's stellar halo (their Figure 9).
For their lowest metallicity bin ($-2.7 < \text{[Fe/H]} < -1.7$), the Auriga density profiles are steeper than that of M31.

The second potential cause is that there are formation mechanisms for horizontal branch stars that are lacking in the AuriDESI stellar evolution models.
In particular, the PARSEC isochrones that we use to generate individual stars assume a constant Reimers mass loss factor $\eta = 0.2$ (see discussion in \citealp{Lee:2022} and more sophisticated treatment in \citealp{Nguyen:2025}) and do not account for binary stellar evolution.
Binary mass loss is one of the main proposed mechanisms for young metal-rich RRLs that are observed in the Milky Way disk \citep{Bobrick:2024, Iorio:2026}, though such stars have not been observed in the halo.
Changing these parameters in the adopted stellar evolution model might enable horizontal branch stars with intermediate ages ($\sim$2-8~Gyr), which preferentially reside in the inner 30~kpc of Auriga galaxies.

\begin{figure*}
    \centering
    \includegraphics[width=\linewidth]{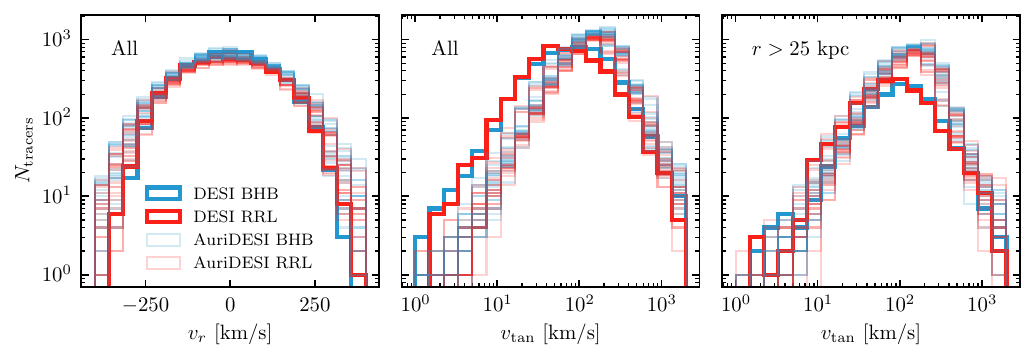}
    \caption{Velocity distributions for BHB and RRL tracers in DESI and AuriDESI (error-convolved), where each AuriDESI mock is represented as a thin line (24 mocks in total).
    The left and center panels show the Galactocentric radial velocity $v_r$ and tangential velocity $v_\mathrm{tan}$, respectively.
    The right panel shows the tangential velocities after selecting for tracers beyond 25~kpc.
    The mock distributions are after all downsampling procedures, including to match the number of tracers in the DESI DR2 catalogues.
    AuriDESI tracers generally reproduce the radial velocity distributions of the DR2 catalogues, but lack a population of nearby stars with low tangential velocities ($v_\mathrm{tan} \lesssim 80$~km~s$^{-1}$).
    }
    \label{fig:velocity-comparison}
\end{figure*}

In Figure~\ref{fig:velocity-comparison} we show the distribution of tracers in Galactocentric radial ($v_r$) and tangential ($v_\mathrm{tan}$) velocity.
The overall $v_r$ distributions for AuriDESI provide a reasonable match to the DR2 tracers, particularly considering that these haloes were not selected to have Milky Way-like accretion histories (in particular, a GSE analogue that produces prominent structures in $r-v_r$ phase space).
The bulk $v_\mathrm{tan}$ distributions peak at slightly higher values than the DR2 data, but with a substantial tail at lower tangential velocities.
As seen in the right-most panel, this low-velocity tail disappears further in the halo ($r > 25$~kpc), bringing the AuriDESI data into better agreement (see \citealp{Deason:2017} for further detail on the bulk rotation of the old stellar halo in Auriga).
We emphasize that the DR2 tracers (in particular BHBs) with high tangential velocities are reproduced in the AuriDESI mocks; such stars have larger distances and larger proper motion uncertainties.
In addition to the comparison shown here, other companion papers analyze the velocity distributions \citep{Zhou:2026empdf} and velocity anisotropy \citep{Valluri:2026nimble} of these mock catalogues, generally finding reasonable agreement with the DR2 data.

In addition to examining the overall phase-space distribution, we also consider whether the AuriDESI tracers have measurement uncertainties that are comparable to the DR2 data.
While the line-of-sight uncertainty models are constructed based on the DR2 tracers (see Figure~\ref{fig:vrad-error-model}), the AuriDESI tracers have flatter distributions in apparent magnitude (i.e., a higher fraction of fainter stars) than their DR2 counterparts.
As a result, the typical uncertainties assigned to the AuriDESI tracers end up being slightly larger.
For DR2 BHBs, the median (90th percentile) uncertainty is 1.65 (6.06)~km~s$^{-1}$, while for RRLs this is 5.87 (46.16)~km~s$^{-1}$.
The corresponding values for AuriDESI BHBs are $\sim$2.89 (10.47)~km~s$^{-1}$, while for AuriDESI RRLs this is 6.87 (51.71)~km~s$^{-1}$.
A similar effect occurs for the proper motion uncertainties as well.

\subsection{Ages and metallicities} \label{sec:age-feh-comparison}

\begin{figure*}
    \centering
    \includegraphics[width=\linewidth]{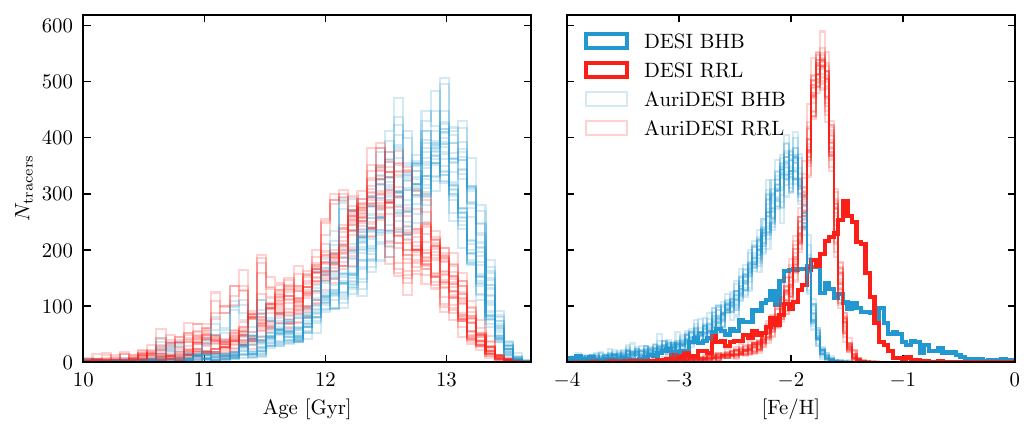}
    \caption{Ages and metallicities for AuriDESI BHBs (blue) and RRLs (red), where each AuriDESI mock is represented as a thin line (24 mocks in total).
    Metallicities for DESI DR2 BHBs and RRLs are also shown with thicker lines.
    The mock distributions are after all downsampling procedures, including to match the number of tracers in the DESI DR2 catalogues.
    AuriDESI BHBs are slightly older and more metal-poor than AuriDESI RRLs, while both are more metal-poor than their corresponding DR2 counterparts.
    }
    \label{fig:age-feh-comparison}
\end{figure*}

In Figure~\ref{fig:age-feh-comparison} we show the ages and metallicities of AuriDESI BHBs and RRLs.
While both populations are old and metal-poor, the BHBs are slightly more so for both properties.
The median age of the AuriDESI BHBs is $\sim$12.6~Gyr and median [Fe/H]~$\sim$~$-2.2$, while the equivalent for AuriDESI RRLs are $\sim$12.4~Gyr and [Fe/H]~$\sim$~$-1.8$.

While age measurements are not available for the DR2 halo stars, we can compare the simulated metallicities to those that are observed.
There is a slight offset in the medians between the two distributions, with DESI BHBs having median [Fe/H]~$\sim$~$-1.9$ and RRLs with [Fe/H]~$\sim$~$-1.7$.
More prominently, both distributions are substantially broader than the AuriDESI versions, particularly for BHBs which have 25~per~cent with [Fe/H] $> -1.5$, while this is effectively the maximum [Fe/H] for AuriDESI BHBs.\footnote{Seven BHBs across all haloes and viewing angles have [Fe/H] $> -1.5$.}

This may be a surprising result, given that satellites and stellar haloes in Auriga are more metal rich than observations \citep{Monachesi:2019, Grand:2021, Thomas:2021, Kizhuprakkat:2024, Riley:2026}.
However, these comparisons are made using full stellar distributions, particularly across all ages.
Old stellar populations that produce horizontal branch stars are more metal-poor in Auriga than the observations.
This suggests that the age-metallicity relation in Auriga may be too steep at very early times (see Section 4.1 of \citealp{Iza:2025} for general discussion on this relation in Auriga), though we caution that this interpretation heavily relies on our adopted PARSEC models.

\subsection{RRL-to-BHB ratio} \label{sec:bhb-rrl-ratio-comparison}

\begin{figure}
    \centering
    \includegraphics[width=\linewidth]{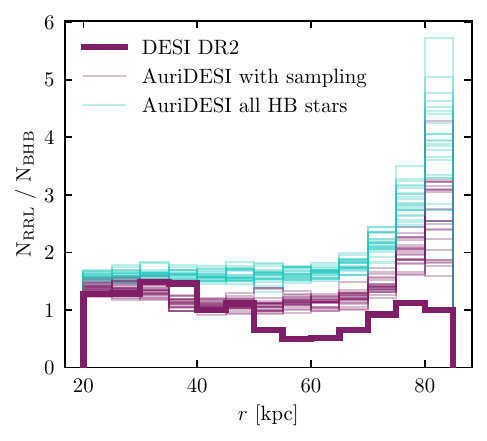}
    \caption{Number of RRLs per BHB as a function of distance, where each AuriDESI mock is represented as a thin line (24 mocks in total).
    The DESI DR2 tracers are shown in solid black, while AuriDESI tracers using all downsampling procedures are shown in red.
    We also show the ratio before applying downsampling procedures (i.e., `photometrically complete') in purple.
    }
    \label{fig:rrl-bhb-ratio}
\end{figure}

We also consider the ratio of RRLs to BHBs, which was reported in \citet{Medina:2025massmodeling} to differ between early versions of these AuriDESI catalogues and the DESI DR1 BHB and RRL catalogues.
In Figure~\ref{fig:rrl-bhb-ratio} we show this ratio $N_\text{RRL} / N_\text{BHB}$ as a function of Galactocentric distance for the DESI DR2 tracers.
There are approximately equal numbers of both tracers out to 50~kpc, beyond which the number of RRLs drops faster than BHBs such that the ratio decreases to $\sim$0.5.
This behaviour is also visible in the overall radial distributions presented in Figure~\ref{fig:distance-comparison}.

The equivalent for the AuriDESI tracers is shown as thin lines.
For our standard AuriDESI sample, which captures all sampling effects described in Section~\ref{sec:tracers}, the RRL-to-BHB ratio is consistent with the DR2 data, with nearly equal numbers in each category.
Unlike the DR2 tracers, the AuriDESI ratio remains roughly equal until 70~kpc, after which BHBs fall faster than RRLs and the ratio increases rapidly.
At the faintest magnitudes that remain within the DESI magnitude selection, the RRL-to-BHB ratio is as high as $\sim$3.

The inclusion or removal of halo substructures (satellite galaxies and massive stellar streams) does not impact this ratio.
However, the typical observational completeness across the sky is higher for BHBs than RRLs (see Section~\ref{sec:desi-selfunc} and left panels of Figure~\ref{fig:spatial-pobs-error}).
If we use all RRL and BHB stars in the DESI footprint, the resulting RRL-to-BHB ratio is shown in purple in Figure~\ref{fig:rrl-bhb-ratio}.
This ratio is always higher than the version that includes all sampling, largely reflecting the difference in completeness of the two samples.

It is not immediately clear whether matching this ratio (or not, depending on the considered radial range) should be considered a success or failure of the AuriDESI mocks.
In addition to the caveats surrounding Auriga and PARSEC that permeate the rest of this section, our ratio of RRL to BHB is also sensitive to the adopted $T_\text{eff}$ separation along the horizontal branch.
As discussed in Section~\ref{sec:tracer-definition}, we adopted a separation boundary at $T_\text{eff} = 7000$~K, where hotter stars are labelled as BHBs and cooler stars are labelled as RRLs.
Calibrating this separation to produce the correct RRL-to-BHB ratio might at first seem a reasonable strategy, but in the case of AuriDESI this would require (a) deciding whether it is more important to match at closer or further distances and (b) some trial and error that must be assessed \textit{after} applying survey selection effects.
Since it is not obvious that this approach would provide any benefits relative to other modeling improvements (e.g.~improved treatment of mass loss for giants), we do not recommend calibrating to reproduce this ratio when constructing future mocks.

\section{Discussion}

\subsection{The Auriga haloes in context}

Although the original sample of Milky Way-mass haloes in the Auriga project was a straightforward selection in halo mass and isolation \citep{Grand:2017}, the six haloes resimulated at higher resolution were selected because of various interesting features of their stellar discs or accretion histories.
Au-6 has disc mass and physical scale lengths that most closely resemble the Milky Way \citep{Bland-Hawthorn:2016, Grand:2018}, along with a very quiet assembly history resulting in the smallest number of overall satellites and stellar streams \citep{Simpson:2018, Riley:2025}.
For this reason, it is generally regarded as the most ``Milky Way-like'' of this sample.
Au-16 and Au-24 were selected due to their relatively massive stellar discs.
Au-23 and Au-27 have experienced close pericentric passages with massive satellites in the past, resulting in perturbed stellar discs.
We caution that the amount of substructure or `disequilibrium' in the stellar halo, and specifically how much it impacts halo mass estimates, is difficult to assess based on a list of accreted satellite masses, pericenters, and merger times.

Despite substantial variations in central disc properties and accretion histories, the resulting mock tracer catalogues are remarkably similar to each other.
The distribution of tracers in terms of distance (Figure~\ref{fig:distance-comparison}), kinematics (Figure~\ref{fig:velocity-comparison}), age and metallicity (Figure~\ref{fig:age-feh-comparison}), and RRL-to-BHB ratio (Figure~\ref{fig:rrl-bhb-ratio}) show relatively little scatter among the various haloes and viewing angles.
This suggests that these distributions are primarily impacted by properties of the overall Auriga and PARSEC models used to generate them, rather than peculiarities of any particular halo.

Finally, we note that the majority of AuriDESI BHBs and RRLs are from accreted particles, with $\sim$15$-$25~per~cent formed in situ.\footnote{The numbers listed in this paragraph are prior to random downsampling to match the number of stars in the DESI DR2 catalogues but include all other effects (footprint, magnitude selection, removing satellites and massive streams).}
The contribution of in situ stars decreases with distance, but is still $\sim$5$-$15~per~cent beyond 50~kpc, in reasonable agreement with the accreted mass fraction profiles reported in \citet{Monachesi:2019}.
Of the accreted stars, most are contributed by a relatively small number of progenitor galaxies.
The number of systems required to contribute over 90~per~cent of accreted tracers in the catalogues ($N_{90}$ in \citealp{Monachesi:2019}) is $\sim$10$-$30.
The weighted average of number of contributing progenitors ($N_\text{sig}$ in \citealp{Cooper:2010, Pu:2025}) is $\sim$4$-$16, though Au-21 is substantially lower ($\sim$4) than the rest ($\sim$8$-$16).
These numbers are much larger than statistics that consider all star particles in the accreted halo \citep{Monachesi:2019, Pu:2025}, suggesting that old and metal-poor stars in the halo are built from many more systems than the overall accreted halo.

\subsection{Limitations and improvements}

A clear result from Section~\ref{sec:dr2-comparison} is that producing a mock survey can incorporate many degrees of realism that are tailored to a particular dataset (e.g., error models and selection functions) but the resulting mocks can still diverge from the real catalogues in key properties.
We emphasize that mock surveys similar to AuriDESI lie at the intersection of three modeling efforts: the underlying cosmological simulations,\footnote{Which are themselves a combination of structure formation (accretion history, dark matter physics) and galaxy formation physics.} the stellar population models, and the implementation of survey properties.
This makes it difficult to trace a mismatch between a mock survey and the data it is emulating to a single cause.

The AuriDESI mock catalogues were designed to support the DR2 mass modeling effort as a validation tool.
In this context, they capture important characteristics and potential biases affecting the DESI data, including the sky coverage and limiting magnitude, measurement uncertainties that are correlated with apparent magnitude, and substructures in the stellar phase-space distribution.
Since these AuriDESI catalogues only produce mock RRL and BHB stars, we encourage users focused on other stellar populations within a DESI-like survey context to use the AuriDESI catalogues presented in \citet{Kizhuprakkat:2024}.
Due to limitations with the underlying Auriga simulations, users should consider other datasets if they are interested in globular clusters, accreted galaxies (regardless of dynamical state) that are below our nominal resolution limit of $M_\star = 10^6$~M$_\odot$, or disequilibrium due to massive satellites that are nearby at the present day (i.e., LMC analogues).

In terms of future improvements for AuriDESI and similar mock efforts within the community, there are two primary routes worth considering.
The first is to implement a variety of underlying stellar population models when generating mocks, either varying parameters within a given model (e.g., varying the Reimers mass loss parameter for PARSEC) or implementing multiple models (e.g., PARSEC, MIST, and BaSTI; \citealp{Bressan:2012, Choi:2016, Hidalgo:2018}).
In particular, some of these models have implemented non-solar [$\alpha$/Fe] ratio isochrones \citep{Fu:2018, Pietrinferni:2021, Dotter:2026}, which could be especially relevant for accreted stellar populations like those studied here.
The second is to generate similarly-designed mock surveys using multiple suites of cosmological simulations.
While both of these suggestions increase the scale of the mock-making effort, they are similar to developments within observations\footnote{For example, measurements of star formation histories can be derived using a variety of theoretical models \citep{McQuinn:2023, McQuinn:2024, Mutlu-Pakdil:2026, Jones:2026}.} and are becoming increasingly feasible with public availability of mock-making tools and simulations.

\subsection{To future mock-makers} \label{sec:lessons}

Producing a set of mock surveys to assess survey systematics and improve measurement quality within a broader collaborative effort is a relatively new workflow for Galactic surveys (though not for large scale structure, e.g., \citealp{Cole:1998, Cole:2005, Percival:2001}).\footnote{We note that the Besan\c{c}on \citep{Robin:2003} and Galaxia \citep{Sharma:2011galaxia} models played an invaluable role in the preparation for \textit{Gaia} \citep{Robin:2012} and DESI \citep{Cooper:2023}, respectively.}
With humility, and the caution that each large collaboration is a unique organism, we share our experience, with the aim of benefiting future efforts of a similar nature.

The first is that mock versions of surveys that are photometrically complete and serve as typical cross-match benchmarks (e.g., \textit{Gaia}) are incredibly useful baselines to select data from.
Adapting these to tailored surveys, as the DESI MWS is broadly selected from \textit{Gaia}, requires close collaboration with those producing the real survey data and more nimble methodologies (e.g., the tailored selection functions described in Section~\ref{sec:desi-selfunc}, or the `spectroscopic' AuriDESI mocks sampled using the DESI fiber assignment algorithm in \citealp{Kizhuprakkat:2024}).

Prior to generating a single mock star, it is worth establishing which properties of the survey are essential and what degree of realism is considered successful.
Using the AuriDESI tracer catalogues as an example, since DESI is designed as a high-latitude low-extinction survey, providing all magnitudes without extinction effects applied was considered reasonable.
However, initial catalogues only included stars brighter than $r=19$ (the bright limit of the main MWS targeting scheme) were viewed as insufficient, since possible tests using fainter stars were envisioned.

Finally, we highlight that tests on mock surveys are typically done before application to the real survey data (or at least contemporaneously).
This is a subtle but important shift from mock surveys being an \textit{output}, tweaked until ready for publication and wider sharing, to being an \textit{input}, versioned and iterated and eventually frozen for downstream users.
Worse still, often such mocks are themselves downstream from other survey outputs (e.g., catalogues, selection functions), such that a change in those outputs may require a new version of mocks.
This means that the course of a project will likely see several major updates, with a final frozen version in place with enough time for downstream users to perform their analysis.

\section{Summary} \label{sec:summary}

In this paper, we present a new set of mock BHB and RRL catalogues that support a broader effort within DESI to measure the mass of the Milky Way.
These mocks are based on the Auriga cosmological simulations \citep{Grand:2017, Grand:2024} and build on the prior mock surveys AuriGaia \citep{Grand:2018} and AuriDESI \citep{Kizhuprakkat:2024}.

Relative to the previous AuriDESI mocks presented in \citet{Kizhuprakkat:2024}, we have refined the selection of horizontal branch stars, updated the treatment of line-of-sight velocity and proper motion uncertainties, and implemented an observational survey selection function that varies across the DESI footprint.
Each of these modifications is tailored specifically to the DESI DR2 BHB and RRL catalogues.
Users are encouraged to use the \citet{Kizhuprakkat:2024} mocks for other stellar populations.

These mock tracer catalogues are used in a series of companion papers \citep{Zhou:2026empdf, Medina:2026gme, Valluri:2026nimble} to validate their techniques to measure the mass of the Milky Way using the DESI DR2 BHB and RRL datasets.
A comparison of their results, joint mass estimate, and other key findings are presented in \citet{DESI:2026mwmass}.
The mock catalogues, along with the true cumulative mass profiles in Auriga, are publicly available on Zenodo and described in Appendix~\ref{app:data-model}.
We encourage modelers using DESI (or sufficiently DESI-like) BHB or RRL catalogues to test their mass estimation methods on these mocks.

\section*{Acknowledgments}

We thank Maddie Lucey, Robyn Sanderson, and Nora Shipp for enlightening discussions.
AHR thanks Taylor Swift's \textit{Red (Taylor's Version)} and Da{\dh}i Freyr's \textit{Too Much Not Enough} for providing a soundscape condusive to writing this manuscript.
This research made extensive use of the Science Explorer, funded by NASA under Cooperative Agreement 80NSSC21M00561, and \href{https://arxiv.org/}{arXiv.org} for bibliographic information.

AHR was supported by a fellowship funded by the Wenner-Gren Foundations (Wenner-Gren Stiftelserna), a Research Fellowship from the Royal Commission for the Exhibition of 1851, and by the Science \& Technology Facilities Council (STFC) through grant ST/T000244/1.
GEM acknowledges support from the Dutch Research Council (NWO) through the Veni grant VI.Veni.252.355.
GEM and TSL acknowledge financial support from Natural Sciences and Engineering Research Council of Canada (NSERC) through grant RGPIN-2022-04794.
APC is supported by a Yushan Fellowship awarded by the Taiwanese Ministry of Education (MOE), MOE-113-YSFMS-0002-001-P2.
APC and NK acknowledge support from Taiwan's National Science and Technology Council (NSTC) under grants 114-2112-M-007-005 and 115-2628-M-007-009-MY4.
SK and TMC acknowledge support from STFC grant ST/Y001001/1.

This work used the DiRAC@Durham facility managed by the Institute for Computational Cosmology on behalf of the STFC DiRAC HPC Facility (\href{https://dirac.ac.uk/}{www.dirac.ac.uk}).
The equipment was funded by BEIS capital funding via STFC capital grants ST/K00042X/1, ST/P002293/1, ST/R002371/1 and ST/S002502/1, Durham University and STFC operations grant ST/R000832/1.
DiRAC is part of the National e-Infrastructure.

For the purpose of open access, the author has applied a Creative Commons Attribution (CC BY) licence to any Author Accepted Manuscript version arising from this submission.

This material is based upon work supported by the U.S.~Department of Energy (DOE), Office of Science, Office of High-Energy Physics, under Contract No.~DE-AC02-05CH11231, and by the National Energy Research Scientific Computing Center, a DOE Office of Science User Facility under the same contract.
Additional support for DESI was provided by the U.S.~National Science Foundation (NSF), Division of Astronomical Sciences under Contract No.~AST-0950945 to the NSF's National Optical-Infrared Astronomy Research Laboratory; the Science and Technology Facilities Council of the United Kingdom; the Gordon and Betty Moore Foundation; the Heising-Simons Foundation; the French Alternative Energies and Atomic Energy Commission (CEA); the Secretariat of Science, Humanities, Technology and Innovation (SECIHTI) of Mexico; the Ministry of Science, Innovation and Universities of Spain (MICIU/AEI/10.13039/501100011033), and by the DESI Member Institutions: \url{https://www.desi.lbl.gov/collaborating-institutions}.
Any opinions, findings, and conclusions or recommendations expressed in this material are those of the author(s) and do not necessarily reflect the views of the U.~S.~National Science Foundation, the U.~S.~Department of Energy, or any of the listed funding agencies.

The authors are honored to be permitted to conduct scientific research on I'oligam Du'ag (Kitt Peak), a mountain with particular significance to the Tohono O'odham Nation.

\section*{Data Availability}
The original AuriDESI mock catalogues \citep{Kizhuprakkat:2024} are publicly available.\footnote{\url{https://data.desi.lbl.gov/public/papers/mws/auridesi/v1/}}
The mock tracer catalogues produced in this work are available on Zenodo: \url{https://doi.org/10.5281/zenodo.23599247}.
We have used simulations from the Auriga Project public data release \citep{Grand:2024} available on the Globus platform.\footnote{\href{https://wwwmpa.mpa-garching.mpg.de/auriga/data.html}{https://wwwmpa.mpa-garching.mpg.de/auriga/data}}

All data shown in the figures, or necessary to reproduce them, are available in a machine-readable form on Zenodo: \url{https://doi.org/10.5281/zenodo.23578705}.

\section*{Software}

This research made use of the Python programming language, along with many community-developed or maintained software packages including:
    \textsc{astropy} \citep{Astropy:2013, Astropy:2018, Astropy:2022},
    \textsc{cmasher} \citep{cmasher, cmasher1.8.0},
    \textsc{cython} \citep{cython},
    \textsc{gala} \citep{gala},
    \textsc{h5py} \citep{h5py, h5py3.7.0},
    \textsc{jupyter} \citep{ipython, jupyter},
    \textsc{matplotlib} \citep{matplotlib},
    \textsc{numpy} \citep{numpy},
    \textsc{pandas} \citep{pandas, pandas1.5.0},
    and \textsc{scipy} \citep{scipy, scipy1.9.1}.
We also thank the maintainers of the \textsc{arepo-snap-util} package and the \textit{Gaia} Project Scientist Support Team and the \textit{Gaia} Data Processing and Analysis Consortium (DPAC).
Parts of the results in this work make use of the colormaps in the \textsc{cmasher} package.
Software citation information aggregated using \href{https://www.tomwagg.com/software-citation-station/}{The Software Citation Station} \citep{software-citation-station-paper, software-citation-station-zenodo}.

Use of large language models (LLMs) in this work was governed by internal DESI Collaboration policy that emphasizes disclosure and ultimate responsibility of the scientist.
AHR used LLM-based tools for inline code suggestions (`autocomplete') and drafting docstrings for prewritten Python code.
Individual authors used LLM-based tools to assist in the following tasks: proofreading manuscript text.
LLMs used include Claude (Anthropic; Opus 5.5) and GPT (OpenAI; GPT-5, GPT-6).
All scientific conclusions are the authors' own.

\bibliographystyle{mnras}

\bibliography{main, software, dr2papers, desipapers}

\begin{appendix}

\section{Data model} \label{app:data-model}
The mock catalogues are in FITS file format and can be downloaded in their entirety from the following Zenodo link: \url{https://doi.org/10.5281/zenodo.23599247}.
Individual mock catalogues are $\sim$10--20 MB, while the full dataset is approximately 600 MB.
The data products and units are listed in Table~\ref{tab:data-model-obs} for observed (i.e., error-convolved) values and Table~\ref{tab:data-model-true} for true values.

\begingroup 
    \setlength{\tabcolsep}{6pt} 
    \renewcommand{\arraystretch}{1.25} 
    \setlength\extrarowheight{2pt}
    \begin{table*}
        \centering
        \begin{tabular}{ l l c}
            \hline \hline
            Quantity & Description & Unit \\
            \hline
            \texttt{obs} HDU (Observed/error-convolved values) \\
            \hline
            \texttt{ra} & Right ascension & deg \\
            \texttt{dec} & Declination & deg \\
            \texttt{distance} & Distance & kpc \\
            \texttt{distance\_err} & Standard error of distance & kpc \\
            \texttt{pmra} & Proper motion in R.A.~direction & mas yr$^{-1}$ \\
            \texttt{pmra\_err} & Standard error of proper motion in R.A.~direction & mas yr$^{-1}$ \\
            \texttt{pmdec} & Proper motion in R.A.~direction & mas yr$^{-1}$ \\
            \texttt{pmdec\_err} & Standard error of proper motion in declination~direction & mas yr$^{-1}$ \\
            \texttt{vrad} (BHB) / \texttt{v0} (RRL) & Line-of-sight velocity & km s$^{-1}$ \\
            \texttt{vrad\_err} (BHB) / \texttt{v0\_err} (RRL) & Standard error of line-of-sight velocity & km s$^{-1}$ \\
            \texttt{parallax} & Parallax & mas \\
            \texttt{parallax\_err} & Standard error of parallax & mas \\
            \texttt{id} & Unique identifier of a tracer (randomly assigned) & -- \\
            \texttt{p\_obs\_loa} & Probability a tracer is observed in DESI DR2 & -- \\
            \texttt{in\_loa} & Boolean flag for one realization of $p_\text{obs}$ & -- \\
            \texttt{id\_loa} & Randomly assigned ID of tracers for which \texttt{in\_dr2} is True & -- \\
            \hline \hline
        \end{tabular}
        \caption{Data model for the \texttt{obs} HDU of the mock catalogues. Note that ``Loa'' is the internal release name for DR2 (e.g., \texttt{p\_obs\_loa}).}
        \label{tab:data-model-obs}
    \end{table*}
\endgroup

\begingroup 
    \setlength{\tabcolsep}{6pt} 
    \renewcommand{\arraystretch}{1.25} 
    \setlength\extrarowheight{2pt}
    \begin{table*}
        \centering
        \begin{tabular}{ l l c}
            \hline \hline
            Quantity & Description & Unit \\
            \hline
            \texttt{true} HDU (True values) \\
            \hline
            \texttt{targetid} & Unique identifier for a tracer & -- \\
            \texttt{ra} & Right ascension & deg \\
            \texttt{dec} & Declination & deg \\
            \texttt{parallax} & Parallax & mas \\
            \texttt{pmra} & Proper motion in R.A.~direction & mas yr$^{-1}$ \\
            \texttt{pmdec} & Proper motion in R.A.~direction & mas yr$^{-1}$ \\
            \texttt{vrad} (both RRL and BHB) & Line-of-sight velocity & km s$^{-1}$ \\
            \texttt{x\_parent} & Galactocentric $X$ coordinate of the parent star particle & kpc \\
            \texttt{y\_parent} & Galactocentric $Y$ coordinate of the parent star particle & kpc \\
            \texttt{z\_parent} & Galactocentric $Z$ coordinate of the parent star particle & kpc \\
            \texttt{vx\_parent} & Velocity of parent star particle along Galactocentric coordinate $X$ & km s$^{-1}$ \\
            \texttt{vy\_parent} & Velocity of parent star particle along Galactocentric coordinate $Y$ & km s$^{-1}$ \\
            \texttt{vz\_parent} & Velocity of parent star particle along Galactocentric coordinate $Z$ & km s$^{-1}$ \\
            \texttt{decam\_g} & Apparent magnitude in DECam $g$-band & mag \\
            \texttt{decam\_r} & Apparent magnitude in DECam $r$-band & mag \\
            \texttt{decam\_z} & Apparent magnitude in DECam $z$-band & mag \\
            \texttt{gaia\_g} & Apparent magnitude in \textit{Gaia} $G$-band & mag \\
            \texttt{gaia\_bp} & Apparent magnitude in \textit{Gaia} $BP$-band & mag \\
            \texttt{gaia\_rp} & Apparent magnitude in \textit{Gaia} $RP$-band & mag \\
            \texttt{age} & Age of parent star particle & Gyr \\
            \texttt{metallicity} & Metallicity of the parent star particle & -- \\
            \texttt{feh} & [Fe/H] abundance of the parent star particle & -- \\
            \texttt{alphafe} & [$\alpha$/Fe] abundance of the parent star particle & -- \\
            \texttt{initial\_mass} & Initial mass & M$_\odot$ \\
            \texttt{mass} & Present-day mass & M$_\odot$ \\
            \texttt{teff} & Stellar effective temperature & K \\
            \texttt{logg} & Surface gravity & log cgs \\
            \texttt{particle\_id} & Unique ID of the parent star particle & -- \\
            \texttt{tree\_id} & ID of progenitor, corresponding to a specific merger tree at the present day & -- \\
            \texttt{subhalo\_nr} & Subhalo the star particle is bound to at the present day & -- \\
            \texttt{pop\_id} & 0: formed \textit{in situ}; 1: has accreted; 2: bound to a satellite at the present day & -- \\
            \texttt{grav\_potential} & Gravitational potential of the parent star particle  & km$^2$ s$^{-2}$ \\
            \texttt{distance} & Heliocentric distance & kpc \\
            \texttt{r\_galcen} & Galactocentric distance & kpc \\
            \texttt{v\_gsr} & Galactic standard of rest velocity & km s$^{-1}$ \\
            \texttt{stream\_flag\_2e9} & Boolean flag if parent star particle is from a stellar stream with $M_\star > 2\times 10^9$~M$_\odot$ & -- \\
            \texttt{stream\_flag\_1e8} & Boolean flag if parent star particle is from a stellar stream with $M_\star > 10^8$~M$_\odot$ & -- \\
            \hline \hline
        \end{tabular}
        \caption{Data model for the \texttt{true} HDU of the mock catalogues.}
        \label{tab:data-model-true}
    \end{table*}
\endgroup

\section{Author Affiliations} \label{app:affiliations}

\input{affiliations}

\end{appendix}

\end{document}

%% file: authors.tex

\def\andname{}

\author{
\vspace{-1.1cm}Alexander~H.~Riley\orcidlink{0000-0001-5805-5766}\altaffilmark{1,2},
Amanda~Bystr\"om\orcidlink{0000-0002-5689-8791}\altaffilmark{3},
Gustavo~E.~Medina\orcidlink{0000-0003-0105-9576}\altaffilmark{4,5},
Namitha~Kizhuprakkat\orcidlink{0000-0003-0853-8887}\altaffilmark{6},
Andrew~P.~Cooper\orcidlink{0000-0001-8274-158X}\altaffilmark{6},
Sergey~E.~Koposov\orcidlink{0000-0003-2644-135X}\altaffilmark{3,7},
T.~M.~Callingham\orcidlink{0000-0003-4952-1299}\altaffilmark{3},
C.~S.~Frenk\orcidlink{0000-0002-2338-716X}\altaffilmark{2},
Ting~S.~Li\orcidlink{0000-0002-9110-6163}\altaffilmark{5},
G.~Thomas\altaffilmark{8},
M.~Valluri\orcidlink{0000-0002-6257-2341}\altaffilmark{9,10},
Y.~Zhou\altaffilmark{11},
J.~Aguilar\altaffilmark{12},
S.~Ahlen\orcidlink{0000-0001-6098-7247}\altaffilmark{13},
F.~Beutler\orcidlink{0000-0003-0467-5438}\altaffilmark{3},
D.~Bianchi\orcidlink{0000-0001-9712-0006}\altaffilmark{14,15},
D.~Brooks\altaffilmark{16},
A.~Carnero Rosell\orcidlink{0000-0003-3044-5150}\altaffilmark{17,8},
T.~Claybaugh\altaffilmark{12},
A.~Cuceu\orcidlink{0000-0002-2169-0595}\altaffilmark{12},
K.~S.~Dawson\orcidlink{0000-0002-0553-3805}\altaffilmark{18},
A.~de la Macorra\orcidlink{0000-0002-1769-1640}\altaffilmark{19},
Arjun~Dey\orcidlink{0000-0002-4928-4003}\altaffilmark{20},
Biprateep~Dey\orcidlink{0000-0002-5665-7912}\altaffilmark{5,21},
P.~Doel\altaffilmark{16},
J.~E.~Forero-Romero\orcidlink{0000-0002-2890-3725}\altaffilmark{22,23},
E.~Gaztañaga\orcidlink{0000-0001-9632-0815}\altaffilmark{24,25,26},
Satya~{Gontcho A Gontcho}\orcidlink{0000-0003-3142-233X}\altaffilmark{27},
G.~Gutierrez\altaffilmark{28},
M.~Ishak\orcidlink{0000-0002-6024-466X}\altaffilmark{29},
J.~Jimenez\orcidlink{0000-0001-8528-3473}\altaffilmark{30},
R.~Kehoe\altaffilmark{31},
A.~Kremin\orcidlink{0000-0001-6356-7424}\altaffilmark{12},
A.~Lambert\altaffilmark{12},
M.~Landriau\orcidlink{0000-0003-1838-8528}\altaffilmark{12},
L.~Le~Guillou\orcidlink{0000-0001-7178-8868}\altaffilmark{32},
A.~Leauthaud\orcidlink{0000-0002-3677-3617}\altaffilmark{33,34},
A.~Meisner\orcidlink{0000-0002-1125-7384}\altaffilmark{20},
R.~Miquel\altaffilmark{35,30},
S.~Nadathur\orcidlink{0000-0001-9070-3102}\altaffilmark{25},
E.~Paillas\orcidlink{0000-0002-4637-2868}\altaffilmark{36},
W.~J.~Percival\orcidlink{0000-0002-0644-5727}\altaffilmark{37,38,39},
F.~Prada\orcidlink{0000-0001-7145-8674}\altaffilmark{40},
I.~P\'erez-R\`afols\orcidlink{0000-0001-6979-0125}\altaffilmark{41},
C.~Ravoux\orcidlink{0000-0002-3500-6635}\altaffilmark{42},
G.~Rossi\altaffilmark{43},
L.~Samushia\orcidlink{0000-0002-1609-5687}\altaffilmark{44,45},
E.~Sanchez\orcidlink{0000-0002-9646-8198}\altaffilmark{46},
C.~Saulder\orcidlink{0000-0002-0408-5633}\altaffilmark{47},
D.~Schlegel\altaffilmark{12},
M.~Schubnell\altaffilmark{48,10},
R.~Sharples\orcidlink{0000-0003-3449-8583}\altaffilmark{49,2},
J.~Silber\orcidlink{0000-0002-3461-0320}\altaffilmark{12},
M.~Siudek\orcidlink{0000-0002-2949-2155}\altaffilmark{26,8},
G.~Tarl\'{e}\orcidlink{0000-0003-1704-0781}\altaffilmark{10},
B.~A.~Weaver\altaffilmark{20},
and R.~Zhou\orcidlink{0000-0001-5381-4372}\altaffilmark{12}
\vspace{1em}
\\ (DESI Collaboration)
\vspace{1em}
}

%% file: affiliations.tex

$^{1}$ Lund Observatory, Division of Astrophysics, Department of Physics, Lund University, SE-221 00 Lund, Sweden\\
$^{2}$ Institute for Computational Cosmology, Department of Physics, Durham University, South Road, Durham DH1 3LE, UK\\
$^{3}$ Institute for Astronomy, University of Edinburgh, Royal Observatory, Blackford Hill, Edinburgh EH9 3HJ, UK\\
$^{4}$ Kapteyn Astronomical Institute, University of Groningen, P.O.Box 880, 9700 AV Groningen, The Netherlands\\
$^{5}$ Department of Astronomy \& Astrophysics, University of Toronto, Toronto, ON M5S 3H4, Canada\\
$^{6}$ Institute of Astronomy and Department of Physics, National Tsing Hua University, 101 Kuang-Fu Rd. Sec. 2, Hsinchu 30013, Taiwan\\
$^{7}$ Institute of Astronomy, University of Cambridge, Madingley Road, Cambridge CB3 0HA, UK\\
$^{8}$ Instituto de Astrof\'{\i}sica de Canarias, C/ V\'{\i}a L\'{a}ctea, s/n, E-38205 La Laguna, Tenerife, Spain\\
$^{9}$ Department of Astronomy, University of Michigan, Ann Arbor, MI 48109, USA\\
$^{10}$ University of Michigan, 500 S. State Street, Ann Arbor, MI 48109, USA\\
$^{11}$ Department of Astronomy, School of Physics and Astronomy, Shanghai Jiao Tong University, Shanghai 200240, China\\
$^{12}$ Lawrence Berkeley National Laboratory, 1 Cyclotron Road, Berkeley, CA 94720, USA\\
$^{13}$ Department of Physics, Boston University, 590 Commonwealth Avenue, Boston, MA 02215 USA\\
$^{14}$ Dipartimento di Fisica ``Aldo Pontremoli'', Universit\`a degli Studi di Milano, Via Celoria 16, I-20133 Milano, Italy\\
$^{15}$ INAF-Osservatorio Astronomico di Brera, Via Brera 28, 20122 Milano, Italy\\
$^{16}$ Department of Physics \& Astronomy, University College London, Gower Street, London, WC1E 6BT, UK\\
$^{17}$ Departamento de Astrof\'{\i}sica, Universidad de La Laguna (ULL), E-38206, La Laguna, Tenerife, Spain\\
$^{18}$ Department of Physics and Astronomy, The University of Utah, 115 South 1400 East, Salt Lake City, UT 84112, USA\\
$^{19}$ Instituto de F\'{\i}sica, Universidad Nacional Aut\'{o}noma de M\'{e}xico,  Circuito de la Investigaci\'{o}n Cient\'{\i}fica, Ciudad Universitaria, Cd. de M\'{e}xico  C.~P.~04510,  M\'{e}xico\\
$^{20}$ NSF NOIRLab, 950 N. Cherry Ave., Tucson, AZ 85719, USA\\
$^{21}$ Department of Physics \& Astronomy and Pittsburgh Particle Physics, Astrophysics, and Cosmology Center (PITT PACC), University of Pittsburgh, 3941 O'Hara Street, Pittsburgh, PA 15260, USA\\
$^{22}$ Departamento de F\'isica, Universidad de los Andes, Cra. 1 No. 18A-10, Edificio Ip, CP 111711, Bogot\'a, Colombia\\
$^{23}$ Observatorio Astron\'omico, Universidad de los Andes, Cra. 1 No. 18A-10, Edificio H, CP 111711 Bogot\'a, Colombia\\
$^{24}$ Institut d'Estudis Espacials de Catalunya (IEEC), c/ Esteve Terradas 1, Edifici RDIT, Campus PMT-UPC, 08860 Castelldefels, Spain\\
$^{25}$ Institute of Cosmology and Gravitation, University of Portsmouth, Dennis Sciama Building, Portsmouth, PO1 3FX, UK\\
$^{26}$ Institute of Space Sciences, ICE-CSIC, Campus UAB, Carrer de Can Magrans s/n, 08913 Bellaterra, Barcelona, Spain\\
$^{27}$ University of Virginia, Department of Astronomy, Charlottesville, VA 22904, USA\\
$^{28}$ Fermi National Accelerator Laboratory, PO Box 500, Batavia, IL 60510, USA\\
$^{29}$ Department of Physics, The University of Texas at Dallas, 800 W. Campbell Rd., Richardson, TX 75080, USA\\
$^{30}$ Institut de F\'{i}sica d’Altes Energies (IFAE), The Barcelona Institute of Science and Technology, Edifici Cn, Campus UAB, 08193, Bellaterra (Barcelona), Spain\\
$^{31}$ Department of Physics, Southern Methodist University, 3215 Daniel Avenue, Dallas, TX 75275, USA\\
$^{32}$ Sorbonne Universit\'{e}, CNRS/IN2P3, Laboratoire de Physique Nucl\'{e}aire et de Hautes Energies (LPNHE), FR-75005 Paris, France\\
$^{33}$ Department of Astronomy and Astrophysics, UCO/Lick Observatory, University of California, 1156 High Street, Santa Cruz, CA 95064, USA\\
$^{34}$ Department of Astronomy and Astrophysics, University of California, Santa Cruz, 1156 High Street, Santa Cruz, CA 95065, USA\\
$^{35}$ Instituci\'{o} Catalana de Recerca i Estudis Avan\c{c}ats, Passeig de Llu\'{\i}s Companys, 23, 08010 Barcelona, Spain\\
$^{36}$ Instituto de Estudios Astrof\'isicos, Facultad de Ingenier\'ia y Ciencias, Universidad Diego Portales, Av. Ej\'ercito Libertador 441, Santiago, Chile\\
$^{37}$ Department of Physics and Astronomy, University of Waterloo, 200 University Ave W, Waterloo, ON N2L 3G1, Canada\\
$^{38}$ Perimeter Institute for Theoretical Physics, 31 Caroline St. North, Waterloo, ON N2L 2Y5, Canada\\
$^{39}$ Waterloo Centre for Astrophysics, University of Waterloo, 200 University Ave W, Waterloo, ON N2L 3G1, Canada\\
$^{40}$ Instituto de Astrof\'{i}sica de Andaluc\'{i}a (CSIC), Glorieta de la Astronom\'{i}a, s/n, E-18008 Granada, Spain\\
$^{41}$ Departament de F\'isica, EEBE, Universitat Polit\`ecnica de Catalunya, c/Eduard Maristany 10, 08930 Barcelona, Spain\\
$^{42}$ Universit\'{e} Clermont-Auvergne, CNRS, LPCA, 63000 Clermont-Ferrand, France\\
$^{43}$ Department of Physics and Astronomy, Sejong University, 209 Neungdong-ro, Gwangjin-gu, Seoul 05006, Republic of Korea\\
$^{44}$ Abastumani Astrophysical Observatory, Tbilisi, GE-0179, Georgia\\
$^{45}$ Department of Physics, Kansas State University, 116 Cardwell Hall, Manhattan, KS 66506, USA\\
$^{46}$ CIEMAT, Avenida Complutense 40, E-28040 Madrid, Spain\\
$^{47}$ Max Planck Institute for Extraterrestrial Physics, Gie\ss enbachstra\ss e 1, 85748 Garching, Germany\\
$^{48}$ Department of Physics, University of Michigan, 450 Church Street, Ann Arbor, MI 48109, USA\\
$^{49}$ Centre for Advanced Instrumentation, Department of Physics, Durham University, South Road, Durham DH1 3LE, UK\\